\documentclass[10pt,onecolumn,aps,prd,preprintnumbers,superscriptaddress,nofootinbib,amsmath,amssymb,floats,floatfix,showkeys,notitlepage,showpacs]{revtex4-1}

\usepackage{orcidlink}
\usepackage{comment}
\usepackage{lipsum}
\usepackage{graphicx}
\usepackage{subfigure}
\usepackage{palatino}
\usepackage{sans}
\usepackage{hyperref}
\hypersetup{colorlinks=true,linkcolor=blue,urlcolor=blue,citecolor=blue}
\usepackage[toc,page]{appendix}
\usepackage[normalem]{ulem}
\usepackage{adjustbox}
\usepackage{latexsym}
\usepackage{amsmath}
\usepackage{amssymb}
\usepackage{amsfonts}
\usepackage{dcolumn}
\usepackage{bm}
\usepackage{tikz}
\usepackage{bigints}
\usepackage{array,tabularx,multirow,booktabs}
\usepackage[tracking=true]{microtype}
\SetTracking{}{500}
\SetTracking{encoding={*}, shape=sc}{40}
\UseRawInputEncoding 
\allowdisplaybreaks

\begin{document}
	\newcommand \nn{\nonumber}
	\newcommand \fc{\frac}
	\newcommand \lt{\left}
	\newcommand \rt{\right}
	\newcommand \pd{\partial}
	\newcommand \e{\text{e}}
	\newcommand \hmn{h_{\mu\nu}}
	\newcommand{\PR}[1]{\ensuremath{\left[#1\right]}} 
	\newcommand{\PC}[1]{\ensuremath{\left(#1\right)}} 
	\newcommand{\PX}[1]{\ensuremath{\left\lbrace#1\right\rbrace}} 
	\newcommand{\BR}[1]{\ensuremath{\left\langle#1\right\vert}} 
	\newcommand{\KT}[1]{\ensuremath{\left\vert#1\right\rangle}} 
	\newcommand{\MD}[1]{\ensuremath{\left\vert#1\right\vert}} 

	
\title{On the analytic generalization of particle deflection in the weak field regime and shadow size in light of EHT constraints for Schwarzschild-like black hole solutions}

\author{Reggie C. Pantig \orcidlink{0000-0002-3101-8591}}
\email{rcpantig@mapua.edu.ph}
\affiliation{Physics Department, Map\'ua University, 658 Muralla St., Intramuros, Manila 1002, Philippines}
	
\begin{abstract}
In this paper, an analytic generalization of the weak field deflection angle (WDA) is derived by utilizing the current non-asymptotically flat generalization of the Gauss-Bonnet theorem. The derived formula is valid for any Schwarzschild-like spacetime, which deviates from the classical Schwarzschild case through some constant parameters. This work provided four examples in the context of bumblebee gravity theory, and one example from a black hole surrounded with soliton dark matter, where some results are new, and some agreed with existing literature. The WDA formula provided a simple calculation, where approximations based on some conditions can be done directly on it, skipping the preliminary steps. For the shadow size analysis, it is shown how it depends solely on the parameter associated with the metric coefficient in the time coordinate. A general formula for the constrained parameter is also derived based on the Event Horizon Collaboration (EHT) observational results. Finally, the work realized further possible generalizations on other black hole models, such as RN-like, dS/AdS-like black hole solutions, and even black hole solutions in higher dimensions.
\end{abstract}	



\maketitle


\section{Introduction}\label{intro}
Black holes are among the most enigmatic and fascinating objects in the universe. Defined by their event horizons, beyond which no information can escape, they challenge our understanding of space, time, and the fundamental laws of physics. Black holes are predicted by Einstein's theory of General Relativity (GR), which describes them as regions where the curvature of spacetime becomes so extreme that not even light can escape. The simplest model is described by the Schwarzschild metric \cite{Schwarzschild:1916uq}. These celestial bodies are not just theoretical constructs; they are astrophysical realities that manifest in various forms, from stellar-mass black holes formed by the collapse of massive stars to the recently discovered supermassive black holes at the centers of galaxies \cite{EventHorizonTelescope:2019dse,EventHorizonTelescope:2019ths, EventHorizonTelescope:2022xqj,EventHorizonTelescope:2022wkp,EventHorizonTelescope:2022wok}.

One of the most profound implications of black hole physics is its impact on gravitational lensing, where light bends as it passes near a massive object such as a black hole. Of particular interest is the weak deflection angle, pioneered by Gibbons and Werner in 2008 for static and spherically symmetric spacetime (SSS), where they used a mathematical tool called the Gauss-Bonnet theorem (GBT) to calculate the deflection angle using the integration of the Gaussian curvature of the corresponding optical metric \cite{Gibbons:2008rj}. While the seminal work successfully calculates the weak deflection angle for asymptotically flat spacetimes, the method fails on non-asymptotically flat spacetimes such as the Kottler spacetime. This problem has been solved through the work of Ishihara et al. \cite{Ishihara:2016vdc} by some change in the integration domain of the Gaussian curvature. That is, by including the finite distance correction of the source and the receiver from the compact object, integrating the Gaussian curvature does not result in divergence. The work is only applied, however, to photons. The formalism for generalization to massive particles and examples can be seen from Refs. \cite{Li:2019vhp,Li:2019qyb,Li:2020wvn,He:2020eah,Li:2020ozr,Li:2021qei,Liu:2022hbp} to cite a few. Then, instead of using the $r\rightarrow \infty$ as part of the integration domain, a more simplified approach was developed by Li et al. \cite{Li:2020wvn} where the photonsphere was used instead. It erases both the ambiguity of the radial distance near the center of the black hole and the remote distance. It only involves the path of the light ray along the photonsphere, and the positions of the source and the receiver. It has been used widely to calculate the weak deflection angles of both null and massive particles \cite{Lambiase:2024uzy,Lambiase:2024vkz,Lobos:2024fzj,Pantig:2024kfn,Mushtaq:2024utq,Liu:2023xtb,Li:2023esz,Mangut:2023oxa,Gao:2023ltr,Huang:2022iwl} to probe parameters that cause deviation from the classical or standard black hole models or various astrophysical contexts. Moreover, observing relativistic image formations due to lensing, it provides precise upper limits on the compactness of massive, dark entities, which is independent of their mass and distance \cite{Virbhadra:1998dy,Virbhadra:1999nm,Virbhadra:2002ju,Virbhadra:2007kw,Virbhadra:2008ws,Adler:2022qtb,Virbhadra:2022ybp,Virbhadra:2022iiy}.

One of the aims of this paper is to derive a general formula for the weak deflection angle, which is valid in a specialized case of Schwarzschild-like metric. Such a metric commonly occurs, for example, in black hole solutions under the bumblebee gravity theory. In classical general relativity, Lorentz symmetry—the principle that the laws of physics are the same for all observers, regardless of their relative velocity or orientation—is a cornerstone. However, in some quantum gravity theories \cite{Kostelecky:1988zi,Colladay:1998fq,Carroll:2001ws,Gambini:1998it}, it's hypothesized that this symmetry might be broken under certain conditions. The Bumblebee model introduces a vector field, commonly referred to as the "Bumblebee field" that is usually denoted as $\ell$, which acquires a nonzero vacuum expectation value. This nonzero value breaks the Lorentz symmetry spontaneously, leading to modified gravitational dynamics. The exact Schwarzschild-like solution in a bumblebee gravity model was derived in Ref. \cite{Casana:2017jkc}, and more extensions to this current model arose and analyzed \cite{Senjaya:2024jgs,Ding:2024qrf,Senjaya:2024ozc,Junior:2024ety,Gogoi:2024scc,AraujoFilho:2024ykw,An:2024fzf,Guo:2023nkd,Lambiase:2023zeo}.

The black hole shadow is another key observational signature that has captured the attention of the scientific community. It is the dark region in the apparent shape of a black hole, surrounded by a bright ring of light formed by photons orbiting close to the event horizon before being bent away by the black hole's gravity. The size and shape of the shadow provide direct information about the geometry of the surrounding spacetime, making it a powerful probe of the underlying gravitational theory. For instance, the Event Horizon Telescope's groundbreaking imaging of the shadow of the supermassive black hole M87* and Sgr. A* marked a significant milestone in black hole astrophysics \cite{EventHorizonTelescope:2019dse,EventHorizonTelescope:2019ths, EventHorizonTelescope:2022xqj,EventHorizonTelescope:2022wkp,EventHorizonTelescope:2022wok}.

In addition to the weak deflection angle formula that will be derived in this paper, a general formula for the shadow radius will also be sought in an attempt to analyze a wide range of black hole models in the Schwarzschild-like solution. The formula will also be connected to the EHT observational constraints to directly and quickly find constraints to the parameter being investigated.

The paper is organized as follows: In Sect. \ref{sec2}, we applied the non-asymptotic generalization of the Gauss-Bonnet theorem to derive the general formula for the weak deflection angle. In Sect. \ref{sec1}, we derived the general formula for the black hole shadow with the EHT constraint parameter. In Sect. \ref{sec3}, we provide examples where these formulas are applied in various Schwarzschild-like black hole solutions. Finally, in Sect. \ref{conc}, final remarks and future research directions are stated. Throughout the paper, $G=c=1$ is used, and the metric signature is $(-,+,+,+)$.

\section{Shadow and analytic constraints of a Schwarzschild-like spacetime} \label{sec1}
Let us consider a spacetime metric that is static and spherically symmetric:
\begin{equation} \label{ds^2}
    {d}{\chi}^{2}=-A(r) {dt}^{2}+B(r){dr}^{2}+C(r) \left({d}\theta^{2}+ \sin ^{2} \theta {d} \phi^{2}\right).
\end{equation}
Without loss of generality, such a metric reduces to a $1+2$ dimensionality as one specializes at $\theta = \pi/2$, leading to
\begin{equation} \label{ds^2_2}
    ds^{2} = -A(r) dt^{2} + B(r) dr^{2} + C(r) d\phi^{2}.
\end{equation}
Next, we assume a Schwarzschild-like configuration of the metric function by introducing some constants $\zeta$ and $\chi$:
\begin{equation} \label{met_func}
    A(r) = \zeta^2 - \frac{2m}{r}, \quad B(r) = \chi^2A(r)^{-1}, \quad C(r) = r^2,
\end{equation}
where $m$ can be characteristically defined as the black hole mass. If one takes $\zeta = \chi = 1$, it is easy to see how the metric reduces to the standard Schwarzschild black hole metric.

Inspecting $A(r)$ in Eq. \eqref{met_func}, one can assign any value for $\zeta$. For instance, if $\zeta = 1$, the Schwarzschild case is recovered if we also assume $\chi = 1$. Otherwise, depending on the expression for $\chi$, we have a certain Schwarzschild-like case such as that in Bumblebee theory where $\chi = \sqrt{1 - \ell}$, where $\ell$ is the Lorentz-violating parameter. If we consider the Newtonian limit to be preserved \cite{Majumder:2024mle} while exploring values of $\zeta$ other than $1$, we can study the approximation $\zeta \to 1$. Such a case implies a small deviation from the Schwarzschild case. It is also important to point out that $\zeta$ may be negative, but such a case will make the metric ill-defined \cite{Majumder:2024mle} if we chose the definition $A(r) = \zeta - 2m/r$. Choosing $\zeta^2$ in $A(r)$ allows negative values and never makes the metric to be ill-defined.

The shadow analysis of any black hole model depends on the null geodesics, that is, on photons' motion around the black hole. The analysis is also less intricate than the derivation of the weak deflection angle. For the complete review of the formalism, see Ref. \cite{Perlick:2021aok}. To derive the black hole shadow radius, we only need the expression for the photonsphere radius, and the critical impact parameter. Following Ref. \cite{Claudel:2000yi,Virbhadra:2002ju,Perlick:2021aok} and using Eq. \eqref{ds^2_2}, the photonsphere radius can be solved via
\begin{equation} 
    A(r)'r^2 - 2A(r)r = 2 \zeta^{2} r -6 m = 0,
\end{equation}
where the notation $\prime$ denotes differentiation with respect to $r$. It simply results to
\begin{equation}
    r_{\rm ph} = \frac{3m}{\zeta^2}.
\end{equation}
The photonsphere radius is seen to be independent of the parameter $\chi$. Using this, the critical impact parameter is found as
\begin{equation}
    b_{\rm crit}^2 = \frac{C(r_{\rm ph})}{A(r_{\rm ph})} = \frac{27m^2}{\zeta^6}.
\end{equation}
Then, we find the exact expression for the radius of the invisible shadow as
\begin{equation} \label{Rsh_ex}
    R_{\rm sh} = b_{\rm crit} \sqrt{A(r_{\rm obs})} = \sqrt{\frac{27 m^{2} }{\zeta^{6}}\left(\zeta^{2}-\frac{2 m}{r_{\rm obs}}\right)}.
\end{equation}
At the location very far from the black hole $r_{\rm obs} \to \infty$, which is applicable to more realistic scenarios,
\begin{equation} \label{Rsh_appx}
    R_{\rm sh} \sim 3 \sqrt{3}\, \frac{m}{\zeta^2} -3 \sqrt{3}\, \frac{m^2}{\zeta^4 r_{\rm obs}} + \mathcal{O}\left(r_{\rm obs}^{-2} \right).
\end{equation}
We should then note that the shadow analysis does not force us to do any approximation in $\zeta$, and shows its independence in $\chi$. Thus, the shadow analysis permits any values for $\zeta$. For instance, setting $\zeta = 1$ above shows the shadow radius for the Schwarzschild black hole. Setting $\zeta = 0$ makes the shadow radius undefined, making any theory about it irrelevant. Interestingly, setting $\zeta < 0$ still allows the formation of the shadow. Lastly, the shadow tends to get bigger as $\zeta$ gets smaller, and vice versa.

The EHT collaboration, as well as other researchers, imposed uncertainties in the shadow radius measurement relative to the Schwarzschild case. At $2\sigma$ level of significance \cite{Vagnozzi:2022moj}, the bounds for Sgr. A* is $4.209M \leq R_{\rm Sch} \leq 5.560M$. As for M87*, it is $ 4.313M \leq R_{\rm Sch} \leq 6.079M$ at $1\sigma$ level \cite{EventHorizonTelescope:2021dqv}. Let $\delta$ represent the difference between upper (or lower) bounds and $R_{\rm Schw}$. Then we can calculate constraint in the parameter $\zeta$ as
\begin{equation} \label{Rsh_cons}
    \zeta = \pm \frac{3^{3/4} \sqrt{m}}{\sqrt{3 \sqrt{3}\, m + \delta}}.
\end{equation}

As we shall see in the next section, the weak deflection angle allows us to constrain $\chi$, a feature that is not present in the black hole shadow analysis.

\section{Weak deflection angle of a Schwarzschild-like spacetime} \label{sec2}
To study the weak deflection angle, we utilize the non-asymptotic version of the Gauss-Bonnet theorem in the calculation of the weak deflection angle \cite{Li:2020wvn}:
\begin{equation} \label{wda_Li}
    \Theta = \iint_{_{r_{\rm ph}}^{\rm R }\square _{r_{\rm ph}}^{\rm S}}KdS + \phi_{\text{RS}},
\end{equation}
to include the effect of the finite distance of the source S and the receiver R. To generalize further, we use the Jacobi metric to include the deflection angle of massive particles of mass $\mu$:
\begin{equation} \label{Jac_met}
    dl^2=g_{ij}dx^{i}dx^{j}
    =(E^2-\mu^2A(r))\left(\frac{B(r)}{A(r)}dr^2+\frac{C(r)}{A(r)}d\phi^2\right).
\end{equation}
Here, $E = (1-v^2)^{1/2}$. It was also shown in Ref. \cite{Li:2020wvn}, using the photonsphere radius $r_{\rm ph}$ as one of the integration domains of the quadrilateral in Eq. \eqref{wda_Li}, that the equation below applies to both massive and null particles:
\begin{equation} \label{wda_Li2}
    \Theta = \int^{\phi_{\rm R}}_{\phi_{\rm S}} \int_{r_{\rm ph}}^{r(\phi)} K\sqrt{g} \, dr \, d\phi + \phi_{\rm RS}.
\end{equation}
From the above equation, $\phi_{\rm RS} = \phi_{\rm R} - \phi_{\rm S}$, where $\phi_{\rm R} = \pi - \phi_{\rm S}$. Furthermore, $K$ is the Gaussian curvature and $g$ is the determinant of the Jacobi metric, defined as follows:
\begin{equation} \label{G_curva}
    K=\frac{1}{\sqrt{g}}\left[\frac{\partial}{\partial\phi}\left(\frac{\sqrt{g}}{g_{rr}}\Gamma_{rr}^{\phi}\right)-\frac{\partial}{\partial r}\left(\frac{\sqrt{g}}{g_{rr}}\Gamma_{r\phi}^{\phi}\right)\right] 
    =-\frac{1}{\sqrt{g}}\left[\frac{\partial}{\partial r}\left(\frac{\sqrt{g}}{g_{rr}}\Gamma_{r\phi}^{\phi}\right)\right],
\end{equation}
and
\begin{equation}
    g = \frac{(E^2 - \mu^2 A(r))B(r)C(r)}{A(r)^2}.
\end{equation}
Due to Eq. \eqref{Jac_met}, we can see that $\Gamma_{rr}^{\phi} = 0$. The consequence of using $r_{\rm ph}$ is then \cite{Li:2020wvn}
\begin{equation}
    \left[\int K\sqrt{g}dr\right]\bigg|_{r=r_{\rm ph}} = 0,
\end{equation}
leading to
\begin{equation} \label{gct}
    \int_{r_{\rm ph}}^{r(\phi)} K\sqrt{g}dr = -\frac{A(r)\left(E^{2}-A(r)\right)C'-E^{2}C(r)A(r)'}{2A(r)\left(E^{2}-A(r)\right)\sqrt{B(r)C(r)}}\bigg|_{r = r(\phi)}.
\end{equation}

The calculation of the integral in Eq. \eqref{gct} requires the orbit equation since due to the upper bound of the integration limit. By setting $u = r^{-1}$,
\begin{equation}
    F(u) \equiv \left(\frac{du}{d\varphi}\right)^2 
    = \frac{C(u)^2u^4}{A(u)B(u)}\Bigg[\left(\frac{E}{J}\right)^2-A(u)\left(\frac{1}{J^2}+\frac{1}{C(u)}\right)\Bigg],
\end{equation}
where $J = Evb$, which is the angular momentum of the particle. Also, note that $b$ is defined as the impact parameter. To reduce clutter, we maintain the variables $E$ and $J$ for now. The above results to
\begin{equation} \label{eorb}
    \left(\frac{du}{d\varphi}\right)^2 = \frac{E^{2}}{\chi^{2} J^{2}}-\frac{\zeta^{2}}{\chi^{2} J^{2}}+\frac{2 m u}{\chi^{2} J^{2}}-\frac{u^{2} \zeta^{2}}{\chi^{2}}+\frac{2 u^{3} m}{\chi^{2}}.
\end{equation}
The next goal is to find the closest approach $u(\phi)$. For a circular orbit, the condition $\left(\frac{du}{d\varphi}\right)^2 = 0$ must apply. It leads to
\begin{equation}
    u = \frac{1}{b}-\frac{\zeta -1}{v^{2} b}.
\end{equation}
Note that the approximation $\zeta \to 1$ is necessary for two reasons. First, we are interested in black holes where the Newtonian limit should be preserved. Second, the analytic expression will become highly complicated without such an approximation. Thus, as a result, it is understood here that $\zeta$ must now take values that are slightly lower or higher than $1$. Proceeding, we know that $u$ is a function of $\phi$, and the equation above does not explicitly show this. We differentiate Eq. \eqref{eorb} again and solve the resulting differential equation, resulting to
\begin{equation}
    u(\phi) = \frac{1}{b}\sin \! \left(\frac{\zeta\,\phi}{\chi} \right)-\frac{\zeta -1}{v^{2} b}.
\end{equation}
We see that $\phi$ in the argument of the sine function is scaled by the ratio of $\zeta$ and $\chi$. Next, we assume a coefficient that is coupled linearly with $m$, say, $p$, and implement the iteration method to complete the function $u(\phi)$. We guess that
\begin{equation}
    u(\phi) = \frac{1}{b}\sin \! \left(\frac{\zeta\,\phi}{\chi} \right)-\frac{\zeta -1}{v^{2} b} + p m,
\end{equation}
and plug this into the orbit equation in Eq. \eqref{eorb}. The result is
\begin{equation} \label{e_u(final)}
    u(\phi) = \frac{1}{b}\sin \! \left(\frac{\zeta\,\phi}{\chi} \right)+ \frac{m}{b^{2} v^{2}}\left[1 + v^2 \cos \! \left(\frac{\zeta\,\phi}{\chi} \right)\right]-\left( \frac{1}{b} + \frac{4m}{b^2}  \right) \frac{c -1}{v^{2}}.
\end{equation}

Using Eq. \eqref{e_u(final)} and going back to Eq. \eqref{gct},
\begin{align}
    &\int_{r_{\rm ph}}^{r(\phi)} K\sqrt{g}dr \sim \frac{\left(2 E^{2}-1\right) m}{\left(E^{2}-1\right) b \chi} \sin \! \left(\frac{\phi}{\chi}\right) -\frac{1}{\chi}-\frac{\zeta -1}{\chi}-\mathcal{O}\left[\left(\zeta-1\right)m \right].
\end{align}
It is surprising how the process naturally removes the parameter $\zeta$ as an argument in the sine function, implying that approximation in $\chi$ may not necessarily be facilitated. Next, we integrate the above resulting to:
\begin{align} \label{e_int}
    &\int_{\phi_{\rm S}}^{\phi_{\rm R}} \int_{r_{\rm ph}}^{r(\phi)} K\sqrt{g} \, dr \, d\phi \sim -\frac{\left(2 E^{2}-1\right) m}{\left(E^{2}-1\right) b}\cos \! \left(\frac{\phi}{\chi}\right) \bigg\vert_{\phi_{\rm S}}^{\phi_{\rm R}} -\frac{\phi_{\rm RS}}{\chi}-\frac{\left(\zeta -1\right) \phi_{\rm RS}}{\chi} + C + \mathcal{O}\left[\left(\zeta-1\right)m \right],
\end{align}
which now require $\phi$ to be solved. Using Eq. \eqref{e_u(final)},
\begin{align} \label{e_phi}
    \phi =  \frac{\chi}{\zeta}\arcsin \! \left(b u \right) + \frac{\chi}{\zeta}\frac{\left[v^{2} \left(b^{2} u^{2}-1\right)-1\right] m}{\sqrt{1-b^{2} u^{2}}\, b \,v^{2}} + \frac{\chi}{\zeta}\frac{\left(\zeta -1\right) }{ v^{2} \sqrt{1-b^{2} u^{2}}} - \mathcal{O}\left[\frac{\chi}{\zeta}\left(\zeta-1\right)m \right].
\end{align}
From here on, we will assume that the finite distance from the black hole of the source and the receiver are equal. It requires us write $\phi_{\rm S} = \varphi$ for brevity. Since the cosine function appeared in Eq. \eqref{e_int},
\begin{align} \label{e_trig}
    \cos \! \left(\frac{\varphi}{\chi}\right) = \sqrt{1-b^{2} u^{2}}-\frac{u \left(v^{2} \left(b^{2} u^{2}-1\right)-1\right) m}{ \sqrt{1-b^{2} u^{2}}\, v^{2}} + \frac{b u \left(\arcsin \! \left(b u \right) v^{2} \sqrt{1-b^{2} u^{2}}-1\right) \left(\zeta -1\right)}{v^{2} \sqrt{1-b^{2} u^{2}}} + \mathcal{O}\left[\left(\zeta-1\right)m \right].
\end{align}
Furthermore, we can apply the following property and definition for $\varphi_{\rm RS}$:
\begin{align}
    \cos \left(\pi - \frac{\varphi}{\chi}\right) = -\cos \left(\frac{\varphi}{\chi}\right), \qquad \phi_{\rm RS} = \pi - 2\varphi.
\end{align}
Using Eq. \eqref{wda_Li2}, the weak deflection angle is derived as
\begin{align} \label{wda_inc}
    \Theta &\sim \frac{2m \left(v^{2}+1\right) \cos \! \left(\frac{\varphi}{\chi} \right)}{v^{2} b} + \left[\frac{\chi}{\zeta}\left(\pi-2\varphi\right)\left(\frac{\zeta}{\chi}-1\right)  \right] + \mathcal{O}\left[\left(\zeta-1\right)m \right].
\end{align}
Note that the equation above is still incomplete. Plugging Eqs. \eqref{e_phi} and \eqref{e_trig} to Eq. \eqref{wda_inc}, we find
\begin{align} \label{wda_gen}
    \Theta &\sim \frac{2m \left(v^{2}+1\right) }{v^{2} b}  \left[ \sqrt{1-b^{2} u^{2}} + \left( \arcsin \! \left(b u \right) - \frac{1}{v^{2} \sqrt{1-b^{2} u^{2}}} \right)bu(\zeta-1) \right] \nn \\
    &- \left[ \pi -2 \arcsin \! \left(b u \right)-\frac{2 \left(\zeta -1\right)}{v^{2} \sqrt{1-b^{2} u^{2}}}-\frac{2m \left[v^{2} \left(b^{2} u^{2}-1\right)-1\right]}{\sqrt{1-b^{2} u^{2}}\, b \,v^{2}} \right]\left( 1-\frac{\chi}{\zeta} \right) - \mathcal{O}\left[\left(\zeta-1\right)m \right].
\end{align}
The equation above is general for any Schwarzschild-like black hole spacetime since it both facilitates the finite-distance correction of the source and receiver, and the deflection angle of massive particles. Note that the finite-distance correction is of tremendous importance if the spacetime is non-asymptotically flat. In the far approximation, however, where $u \rightarrow 0$, Eq. \eqref{wda_gen} reduces to
\begin{align} \label{wda_time}
    \Theta^{\rm massive} &\sim \frac{2m\left(v^{2}+1\right)}{bv^{2}} - \left\{\pi - \frac{2}{v^2}\left[ \zeta-1 - \frac{m(v^2+1)}{b}  \right] \right\}\left( 1-\frac{\chi}{\zeta} \right) + \mathcal{O}\left[\left(\zeta-1\right)m \right].
\end{align}
For massless particles such as photons, where $v = 1$, we obtain the following simplified expression:
\begin{align} \label{wda_null}
    \Theta^{\rm photon} \sim \frac{4m}{b}-\left[\pi - 2\left( \zeta-1 - \frac{2m}{b}  \right) \right]\left( 1-\frac{\chi}{\zeta} \right) + \mathcal{O}\left[\left(\zeta-1\right)m \right].
\end{align}
It is easy to see how this formula will reduce to the Schwarzschild case when $\zeta = \chi = 1$.

\subsection{Solar System Test}
In the weak field regime, it is possible to use the solar system test to constrain parameters coming from the weak deflection angle expression such as in Eq. \eqref{wda_null}. In the parametrized post-Newtonian formalism (PPN), the deflection angle of light reads \cite{Chen:2023bao}
\begin{equation} \label{ppn}
    \Theta^{\rm PPN} \backsimeq \frac{4M_{\odot}}{R_{\odot}}\left(\frac{n \pm \Delta}{2} \right),
\end{equation}
where $n = 1.9998$, and $\Delta = 0.0003$ \cite{fomalont2009progress}. Here, $\Delta$ represents the uncertainty found on the curvature of spacetime caused by the Sun's gravity ($M_\odot = 1476.61 \text{ m}$, $R_{\odot} = 6.96\times 10^{8} \text{ m}$). Note that $\Theta^{\rm PPN}$ is given in radians, and is linked to solar system observations of light bending around the Sun, especially from astrometric measurements like the Very Long Baseline Array (VLBA) \cite{fomalont2009progress}. Comparison of Eq. \eqref{ppn} and Eq. \eqref{wda_null}, we can yield some constraints in $\chi$ given as
\begin{equation} \label{wda_cons}
    \chi \sim \zeta + \frac{2 M_{\odot} \left(n +\Delta -2\right) \left[\left(\pi +2\right) \zeta - 1 \right]}{\pi^{2} R_{\odot}} + \mathcal{O}(M_{\odot}^2).
\end{equation}
Even if we use the solar system test with consideration for the Sun as the compact object, take note that Eq. \eqref{wda_cons} is also valid for supermassive black holes as long as the observational parameter $(n \pm \Delta)$ is given.

\subsection{Einstein ring}
The Einstein ring, denoted by $\theta_{\rm E}$ measured in $\mu\text{as}$, is an observable associated with the weak deflection angle. The position of the weak field images under the thin lens approximation is found as \cite{Bozza:2008ev}
\begin{equation}
\mathcal{D}_\text{RS}\tan\beta = \frac{\mathcal{D}_\text{R}\sin\theta_{\rm E}-\mathcal{D}_\text{S}\sin(\Theta^{\rm photon}-\theta_{\rm E})}{\cos(\Theta^{\rm photon}-\theta_{\rm E})},
\end{equation}
where $\mathcal{D}_\text{S}$ and $\mathcal{D}_\text{R}$ are the distance of the source and the receiver respectively from the lensing object that is assumed to be far away and equidistant. Then, $D_\text{RS}=D_\text{R}+D_\text{S}$. The condition $\beta = 0$ enforces the formation of the Einstein ring:
\begin{equation}
    \theta_{\rm E} = \frac{\mathcal{D}_\text{S}}{\mathcal{D}_\text{RS}}\Theta^{\rm photon}.
\end{equation}
Furthermore, under the assumption that the Einstein ring is very small, then the impact parameter can be approximated as $b \sim \mathcal{D}_\text{R}\sin\theta_{\rm E} \sim \mathcal{D}_\text{R}\theta_{\rm E}$. Assuming $\mathcal{D}_\text{S} = \mathcal{D}_\text{R}$, $\mathcal{D}_\text{RS} = 2 \mathcal{D}_\text{R}$, and $\mathcal{D}_\text{R} \to \infty$, we find
\begin{equation} \label{Ering}
    \theta_{\rm E} \sim \frac{4m}{\left[\pi +2(1-\zeta)\right] \left(\zeta/\chi -1 \right) \mathcal{D}_\text{R}} -\mathcal{O}(\mathcal{D}_\text{R}^{-2}).
\end{equation}

Similar to PPN formalism, let us assume that there is a constrain for $\theta_{\rm E}$:
\begin{equation} \label{Ering_cons}
    \theta_{\rm E} \sim \sqrt{\frac{2m}{\mathcal{D}_\text{R}}}\mathcal{C},
\end{equation}
where $\mathcal{C}$ related to some uncertainty found from some astronomical observations. Then, comparing the above equation with Eq. \eqref{Ering}, one could find constraints in $\chi$:
\begin{equation}
    \chi \sim \zeta -\frac{2 \left[\left(\pi +2\right) \zeta - 2 \right]}{\pi^{2} \mathcal{C}} \sqrt{\frac{2m}{\mathcal{D}_\text{R}}} + \mathcal{O}(\mathcal{D}_\text{R}^{-1}).
\end{equation}
We remark that the above formula is only useful if both $\mathcal{C}$ and $\zeta$ are known, where $\zeta \to 1$ is carried out from some weak field test for black holes.

In the next section, we will apply the derived equations to some examples of Schwarzschild-like black hole solutions. The most prominent is the application of the bumblebee gravity to obtain the Schwarzschild-like metric. Also, some dark matter models, such as that of solitonic dark matter, yield a Schwarzschild-like black hole solution.

\section{Examples in Schwarzschild-like black hole solutions} \label{sec3}
\subsection{Example 1}
In the Schwarzschild-like black hole in the bumblebee gravity model \cite{Casana:2017jkc}, the metric functions are
\begin{equation}
    A(r) = 1 - \frac{2m}{r}, \quad B(r) = (1+\ell)A(r)^{-1}, \quad C(r) = r^2,
\end{equation}
where $\ell$ is the Lorentz violating parameter. Thus, we can write $\zeta = 1$, and $\chi=\sqrt{1+\ell}$. For the shadow, since $\zeta=1$ and the shadow radius is independent of $\chi$. Therefore, $R_{\rm sh} = R_{\rm Schw}$, and we cannot probe the Lorentz-violating parameter using the EHT constraints.

It then indicates how convenient the weak deflection angle is in probing parameters restricted in the shadow analysis. Using Eq. \eqref{wda_gen} gives
\begin{align} \label{wda_gen_case1}
    \Theta = \frac{2m \left(v^{2}+1\right) }{v^{2} b}  \sqrt{1-b^{2} u^{2}} - \left\{ \pi -2 \arcsin \! \left(b u \right)-\frac{2m \left[(-1+v^{2} \left(b^{2} u^{2}-1\right)\right]}{\sqrt{1-b^{2} u^{2}}\, b \,v^{2}} \right\}\left( 1-\sqrt{1+\ell} \right). 
\end{align}
In the approximation $u\rightarrow 0$, the massive particle deflection angle is found as
\begin{align} \label{wda_time_case1}
    \Theta^{\rm massive} = \frac{2m \left(v^{2}+1\right) }{v^{2} b} - \left[ \pi +\frac{2 m \left(v^{2}+1\right)}{b \,v^{2}} \right]\left( 1-\sqrt{1+\ell} \right).
\end{align}
For photons where $v = 1$, we obtain
\begin{align} \label{wda_null_case1}
    \Theta^{\rm photon} &= \frac{4m}{b}-\left( \pi +\frac{4m}{b} \right)\left( 1- \sqrt{1+\ell} \right) \nn \\
    & = \pi\left(\sqrt{1+\ell}-1\right)+ \frac{4m\sqrt{1+\ell}}{b},
\end{align}
which is consistent with what was found in \cite{Li:2020dln,Li:2020wvn,DCarvalho:2021zpf}. We should note that we do not need to implement the approximation $\ell \rightarrow 0$. Next, implementing solar system test through Eq. \eqref{wda_cons}, we find
\begin{equation} \label{e1_cons}
    \ell \sim \frac{4 \left(n +\Delta -2\right) M_\odot}{\pi  R_\odot} + \mathcal{O}(R_\odot^{-2}).
\end{equation}
Hence, the Lorentz-violating parameter is constrained through the bounds
\begin{equation} \label{e1_cons2}
    -1.35 \times 10^{-9} \leq \ell \leq 2.70 \times 10^{-10}.
\end{equation}

\subsection{Example 2}
Next, we consider the Schwarzschild-like black hole in the bumblebee gravity model with the Kalb-Ramond field (or KR field) \cite{Duan:2023gng}:
\begin{equation}
    A(r) = \frac{1}{1-\ell} - \frac{2m}{r}, \quad B(r) = A(r)^{-1}, \quad C(r) = r^2.
\end{equation}
Again, $\ell$ is the Lorentz-violating parameter. We can then write $\zeta=1/\sqrt{1-\ell}$, and $\chi = 1$. We can expect that $\ell$ can be constrained through the shadow analysis. Using Eq. \eqref{Rsh_ex}, it is given exactly as
\begin{equation}
    R_{\rm sh} = \frac{3\sqrt{3}m}{(1-\ell)^{3/2}}\sqrt{\left( \frac{1}{1-\ell}- \frac{2m}{r_{\rm obs}} \right)}.
\end{equation}
In the far approximation, where $r_{\rm obs} \rightarrow \infty$,
\begin{equation}
    R_{\rm sh} \sim 3 \sqrt{3}\, m \left(1-\ell  \right) + \frac{3 \sqrt{3}\, m^{2} \left(1-\ell \right)^{2}}{k} - \mathcal{O}(r_{\rm obs}^{-2}).
\end{equation}
We find the constraint in $\ell$ as
\begin{equation}
    \ell = - \frac{\delta \sqrt{3}}{9m}.
\end{equation}
For M87*, where $\delta = \pm 0.883 m$, $\ell$ is bounded as $-0.17 \leq \ell \leq 0.17$. For Sgr. A*, $-0.364 \leq \delta \leq 0.987$. Hence, $-0.190 \leq \ell \leq 0.070$.

Next, we analyze the weak deflection angle. In this case, however, we require the approximation $\ell \rightarrow 0$. The result for Eq. \eqref{wda_gen} is
\begin{align} \label{wda_gen_case2}
    \Theta &\sim \frac{2m \left(v^{2}+1\right) }{v^{2} b} \left( \sqrt{1-b^{2} u^{2}} \right) \nn \\ 
    &+ \left\{ \arcsin \! \left(b u \right) - \frac{\pi}{2} + \frac{\left(-1+v^{2} \left(b^{2} u^{2}-1\right)\right) m}{\sqrt{-b^{2} u^{2}+1}\, b \,v^{2}} + \frac{\left(v^{2}+1\right) m u}{v^{2}} \left( \arcsin \! \left(b u \right)-\frac{1}{v^{2} \sqrt{-b^{2} u^{2}+1}}  \right)\right\}\ell + \mathcal{O}(\ell^2).
\end{align}
In the far approximation, we obtained
\begin{align} \label{wda_time_case2}
    \Theta^{\rm massive} &\sim \frac{2m \left(v^{2}+1\right) }{v^{2} b} - \left( \frac{\pi}{2} + \frac{(1+v^2)m}{bv^2}  \right)\ell + \mathcal{O}(\ell^2),
\end{align}
and for the case of deflection angle due to photons,
\begin{align} \label{wda_null_case2}
    \Theta^{\rm photon} \sim \frac{4m}{b} - \frac{\pi \ell}{b} - \frac{2m \ell}{b} + \mathcal{O}(\ell^2).
\end{align}
Finally, we constrain $\ell$ using the solar system test. Interestingly, using Eq. \eqref{wda_cons} and doing the necessary approximations, we find
\begin{equation} \label{e2_cons}
    \ell \sim -\frac{4 \left(n +\Delta -2\right) M_\odot}{\pi  R_\odot} - \mathcal{O}(R_\odot^{-2}).
\end{equation}
We see that the only difference of the above expression to Eq. \eqref{e1_cons} is the sign. Hence, the Lorentz-violating parameter is constrained in the same way as in Eq. \eqref{e1_cons2}.

\subsection{Example 3}
In a bumblebee black hole with a global monopole, the metric functions are \cite{Lin:2023foj}
\begin{equation}
    A(r) = 1 - 8\pi \eta^2 - \frac{2m}{r}, \quad B(r) = (1+L) A(r)^{-1}, \quad C(r) = r^2,
\end{equation}
where $L$ is the LSB parameter, and $\eta$ is the parameter related to the global monopole. For the shadow analysis, we expect that it will not depend on the parameter $L$. The exact expression is
\begin{equation}
    R_{\rm sh} = \frac{3 \sqrt{3}\, m}{\left(1-8 \pi  \,\eta^{2}\right)^{3/2}}\sqrt{1-8 \pi  \,\eta^{2}-\frac{2 m}{r_{\rm obs}}}.
\end{equation}
When $r_{\rm obs} \rightarrow \infty$, the expression becomes
\begin{equation}
    R_{\rm sh} \sim \frac{3\sqrt{3}m}{1-8\pi\eta^2}- \frac{3\sqrt{3}m^2}{(1-8\pi\eta^2)^2r_{\rm obs}}  - \mathcal{O}(r_{\rm obs}^{-2}).
\end{equation}
Then, constraint in the weak field limit is found using Eq. \eqref{Rsh_cons}:
\begin{equation}
    \eta = \pm \frac{1}{4}\sqrt{\frac{2\delta}{\pi(3\sqrt{3}m+\delta)}}.
\end{equation}
Using the bounds for $\delta$, the bounds on $\eta$ for M87* are $0.076 \leq \eta \leq 0.090$ and for Sgr. A*, $0.055 \leq \eta \leq 0.080$.

Next, we derive the formula for the weak deflection angle where one should note the approximation $\zeta \rightarrow 1$. 
\begin{align} \label{wda_gen_case3}
    \Theta &= \frac{2m \left(v^{2}+1\right) }{v^{2} b}  \left( \sqrt{1-b^{2} u^{2}} \right) - \left\{ \pi -2 \arcsin \! \left(b u \right)-\frac{2 \left(-1+v^{2} \left(b^{2} u^{2}-1\right)\right) m}{\sqrt{-b^{2} u^{2}+1}\, b \,v^{2}} \right\} \left( 1 - \sqrt{1+L} \right) \nn \\
    & + 4\pi\left\{ \pi -2 \arcsin \! \left(b u \right)-\frac{2 \left(-1+v^{2} \left(b^{2} u^{2}-1\right)\right) m}{\sqrt{-b^{2} u^{2}+1}\, b \,v^{2}} \right\} \sqrt{1+L} \eta^2 - \frac{8 \pi \eta^2  \left(1-\sqrt{1+L}\right)}{v^{2} \sqrt{-b^{2} u^{2}+1}} \nn \\
    & - \frac{8 \pi  \left(v^{2}+1\right) m u \,\eta^{2}}{v^{2}} \left[ \arcsin \! \left(b u \right)-\frac{1}{v^{2} \sqrt{-b^{2} u^{2}+1}} \right] + \mathcal{O}(\eta^4).
\end{align}
In the far approximation, we obtained
\begin{align} \label{wda_time_case3}
    \Theta^{\rm massive} &\sim \frac{2m \left(v^{2}+1\right) }{v^{2} b} - \left[\pi +\frac{2 \left(v^{2}+1\right) m}{b \,v^{2}}\right] \left(1-\sqrt{1+L}\right) \nn \\
    &+ \left\{4 \left[\pi +\frac{2 \left(v^{2}+1\right) m}{b \,v^{2}}\right] \pi  \sqrt{1+L}-\frac{8 \pi  \left(1-\sqrt{1+L}\right)}{v^{2}} \right\} \eta^2 + \mathcal{O}(\eta^4),
\end{align}
and for the case of deflection angle due to photons,
\begin{align} \label{wda_null_case3}
    \Theta^{\rm photon} \sim \pi\left(\sqrt{1-L}-1\right)+ \frac{4m\sqrt{1-L}}{b} + 4 \pi  \,\eta^{2} \left[\pi\sqrt{1+L} +2\left( \sqrt{1+L}-1\right)\right] + \frac{16 m \eta^{2} \pi  \sqrt{1+L}}{b} + \mathcal{O}(\eta^4),
\end{align}
which is quite a worked-out equation. Here, we see that the form of the first two terms is similar to that found in Eq. \eqref{wda_null_case1}.

Now, we find constraints for $L$ using the solar system test. We found through the strong field analysis (through the shadow constraints) that $\eta$ is close to zero, which may indicate that it will be vanishingly small in the weak field regime. Hence, to simplify the constraints through the solar system test, we set $\eta = 0$. One can verify that the result would be the same as Eq. \eqref{e1_cons}, giving the same bounds in eq. \eqref{e1_cons2}.

\subsection{Example 4}
In an Einstein-Hilbert-Bumblebee (EHB) gravity around global monopole field $\bar{\mu}$, the Schwarzschild-like black hole solution is described by the following metric coefficients \cite{Gullu:2020qzu}:
\begin{equation}
    A(r) = 1 - \bar{\mu} - \frac{2m}{r}, \quad B(r) = (1+\ell) A(r)^{-1}, \quad C(r) = r^2.
\end{equation}
We could then write $\zeta = \sqrt{1 - \bar{\mu}}$, and $\chi = \sqrt{1+\ell}$ as usual. For this model then, the shadow analysis gives the exact shadow radius:
\begin{equation}
    R_{\rm sh} = \frac{3 \sqrt{3}\, m}{\left(1-\bar{\mu}\right)^{3/2}}\sqrt{1-\bar{\mu}-\frac{2 m}{r_{\rm obs}}},
\end{equation}
showing independence in the parameter $\ell$. In the far approximation,
\begin{equation}
    R_{\rm sh} \sim \frac{3\sqrt{3}m}{1-\bar{\mu}}- \frac{3\sqrt{3}m^2}{(1-\bar{\mu})^2r_{\rm obs}}  - \mathcal{O}(r_{\rm obs}^{-2}),
\end{equation}
and with the EHT constraints,
\begin{equation}
    \bar{\mu} = \frac{\delta}{3\sqrt{3}m + \delta}.
\end{equation}
With the values $\delta$ assigned for M87* and Sgr. A*, the bounds for $\bar{\mu}$ are $-0.205 \leq \bar{\mu} \leq 0.145$ and $-0.075 \leq \bar{\mu} \leq 0.160$, respectively.

We then derived the weak deflection angle with finite-distance correction as
\begin{align} \label{wda_gen_case4}
    \Theta &\sim \frac{2m \left(v^{2}+1\right) }{v^{2} b}  \left( \sqrt{1-b^{2} u^{2}} \right) - \left\{ \pi -2 \arcsin \! \left(b u \right)-\frac{2 \left(-1+v^{2} \left(b^{2} u^{2}-1\right)\right) m}{\sqrt{-b^{2} u^{2}+1}\, b \,v^{2}} \right\} \left( 1 - \sqrt{1+\ell} \right) \nn \\
    & + \left\{ \pi -2 \arcsin \! \left(b u \right)-\frac{2 \left(-1+v^{2} \left(b^{2} u^{2}-1\right)\right) m}{\sqrt{-b^{2} u^{2}+1}\, b \,v^{2}} \right\} \frac{\bar{\mu}\sqrt{1+\ell}}{2} - \frac{\bar{\mu}  \left(1-\sqrt{1+\ell}\right)}{v^{2} \sqrt{-b^{2} u^{2}+1}} \nn \\
    & - \frac{\left(v^{2}+1\right) m u \bar{\mu} }{v^{2}} \left[ \arcsin \! \left(b u \right)-\frac{1}{v^{2} \sqrt{-b^{2} u^{2}+1}} \right] + \mathcal{O}(\bar{\mu}^2).
\end{align}
In the approximation $u\rightarrow 0$, the massive particle deflection angle is
\begin{align} \label{wda_time_case4}
    \Theta^{\rm massive} &\sim \frac{2m \left(v^{2}+1\right) }{v^{2} b} - \left[ \pi +\frac{2 m \left(v^{2}+1\right)}{b \,v^{2}} \right]\left( 1-\sqrt{1+\ell} \right) \nn \\
    &+  \left\{ \frac{\sqrt{1+\ell}}{2} \left[ \pi +\frac{2 m \left(v^{2}+1\right)}{b \,v^{2}} \right] - \frac{1 - \sqrt{1+\ell}}{v^2} \right\}\bar{\mu} + \mathcal{O}(\bar{\mu}^2).
\end{align}
For photons where $v = 1$, we obtain
\begin{align} \label{wda_null_case4}
    \Theta^{\rm photon} \sim \pi\left(\sqrt{1+\ell}-1\right)+ \frac{4m\sqrt{1+\ell}}{b} + \bar{\mu}\sqrt{1+\ell}\left[\left( \frac{\pi}{2} + \frac{2m}{b} +1 \right) - 1  \right] + \mathcal{O}(\bar{\mu}^2).
\end{align}
Note, how it again retrieves the original first two terms from the bumble bee BH solution, and obtains the correction due to the topological defect parameter $\bar{\mu}$. With the same arguments for constraining $\ell$ as in Example 3, it can be concluded that we should find the same bounds for $\ell$ as given in Eq. \eqref{e1_cons2}.

\subsection{Example 5}
In Ref. \cite{Pantig:2022sjb}, the metric functions of a black hole surrounded with dark matter from the quantum wave model (or solitonic dark matter) are
\begin{equation}
    A(r) = 1 - \frac{4\pi k}{7\alpha r_{c}} - \frac{2m}{r}, \quad B(r) = A(r)^{-1}, \quad C(r) = r^2.
\end{equation}
Here, $k$ is the mass parameter of the solitonic dark matter (as a fraction of the black hole mass), $\alpha = \sqrt[8]{2}-1 \sim 0.09051$ is a constant derived from the half-density comoving core radius $r_{c}$. Thereby, we write $\zeta = \sqrt{1 - \frac{4\pi k}{7\alpha r_{c}}}$, and $\chi = 1$. Under the influence of the solitonic dark matter, the exact expression for the shadow radius is
\begin{equation}
    R_{\rm sh} = \frac{3 \sqrt{3}\, m}{\left(1-\frac{4\pi k}{7\alpha r_{\rm c}}\right)^{3/2}}\sqrt{1-\frac{4\pi k}{7\alpha r_{\rm c}}-\frac{2 m}{r_{\rm obs}}}.
\end{equation}
In the approximation $r_{\rm obs} \rightarrow \infty$,
\begin{equation}
    R_{\rm sh} \sim \frac{3\sqrt{3}m}{1-\frac{4\pi k}{7\alpha r_{\rm c}}}- \frac{3\sqrt{3}m^2}{\left(1-\frac{4\pi k}{7\alpha r_{\rm c}}\right)^2r_{\rm obs}}  - \mathcal{O}(r_{\rm obs}^{-2}),
\end{equation}
and the constraint to the ratio of dark matter mass $k$ and solitonic core $r_{\rm c}$ is given as
\begin{equation}
    \frac{k}{r_{\rm c}} = \frac{7\alpha\delta}{4\pi(3\sqrt{3}m + \delta)}.
\end{equation}
Then, we should find for M87* and Sgr. A* that $\bar{\mu}$ are $-0.010 \leq \bar{\mu} \leq 0.007$ and $-0.004 \leq \bar{\mu} \leq 0.008$, respectively. Note that the upper bounds must be chosen for physical standpoint.

Next, we examine the weak deflection angle. Eq. \eqref{wda_gen} then reduces to
\begin{align} \label{wda_gen_case5}
    \Theta &\sim \frac{2m \left(v^{2}+1\right) }{v^{2} b}  \left( \sqrt{1-b^{2} u^{2}} \right) + \left\{ \pi -2 \arcsin \! \left(b u \right)-\frac{2 \left(-1+v^{2} \left(b^{2} u^{2}-1\right)\right) m}{\sqrt{-b^{2} u^{2}+1}\, b \,v^{2}} \right\} \frac{2\pi k}{7\alpha r_{\rm c}} \nn \\
    & - \frac{4\pi m u k \left(v^{2}+1\right)}{7\alpha r_{\rm c} v^{2}} \left[ \arcsin \! \left(b u \right)-\frac{1}{v^{2} \sqrt{-b^{2} u^{2}+1}} \right] + \mathcal{O}(\bar{\mu}^2).
\end{align}
In the far approximation,
\begin{align} \label{wda_time_case5}
    \Theta^{\rm massive} &\sim \frac{2m \left(v^{2}+1\right) }{v^{2} b} + \frac{2 \pi  k}{7 \alpha  r_{c}}\left[ \frac{2m(1+v^2)}{bv^2} \right] + \mathcal{O}(k^2),
\end{align}
and for the case of deflection angle due to photons,
\begin{align} \label{wda_null_case5}
    \Theta^{\rm photon} \sim \frac{4m}{b} + \frac{2 \pi^{2} k}{7 \alpha  r_{c}} + \frac{8 \pi  k m}{7 \alpha  r_{c} b} + + \mathcal{O}(k^2)
\end{align}
It is consistent, and a more concise version of the expression found in \cite{Pantig:2022sjb}, where it is seen that dark matter effects increase the weak deflection angle. Finally, we can use the solar system test to potentially detect the solitonic dark matter. We find that
\begin{equation}
    \frac{k}{r_{\rm c}} \sim \frac{7 \left(n +\Delta -2\right) M_\odot \alpha}{\pi^2  R_\odot} + \mathcal{O}(R_\odot^{-2}),
\end{equation}
which gives the bounds $-6.81 \times 10^{-11} \leq k/r_{\rm c} \leq 1.36 \times 10^{-11}.$ It indicates that the dark matter content in the solar system is so low, as compared to the upper bounds found for black holes.

\section{Conclusions} \label{conc}
It is without a doubt that the non-asymptotic generalization of the GBT using the photonsphere as part of its integration domain is beautiful in its own right. This work has shown its utility by deriving a general formula for the weak deflection angle, which is valid for a specific class of black hole solution - the Schwarzschild-like solution. Such a remarkable example is the black hole solution due to the bumblebee gravity model, and some cases of solution for black holes surrounded with dark matter. Furthermore, a general formula for the shadow radius with EHT constraints was added, adding more to the scope of the study. Five examples were shown that directly use the derived formulas, which skip the preliminary calculations such as that of the Gaussian curvature, the orbit equation, or the separation angle. Thus, the formula is valid for any recognizable Schwarzschild-like solution as long as the parameters $\zeta$ and $\chi$ are constants and coordinate independent. It is also important to remark that to this date, no deflection angle formula is expressed directly in terms of the metric functions.

As a final remark, it became clear in the derivation of the weak field deflection how it forces the approximation of the parameter $\zeta$, but not necessarily on $\chi$. The latter is also shown to be independent of the shadow radius calculation. In the calculation of the weak deflection angle, the approximation on $\chi$ is necessary if the parameter where it depends is the same as $\zeta$. It is also interesting how the parameter associated with $\zeta$ is automatically approximated to the weak field regime as one uses the bounds of uncertainties for the shadow radius as reported by the EHT collaboration.

There are several research directions based on the results presented in this paper. These are possible extensions to $(1)$ Schwarzschild-like solution in higher dimensional case, $(2)$ the stationary axisymmetric (SAS) case, and $(3)$ solutions that involve black hole charge $Q$ (RN-like solutions) and the cosmological constant $\Lambda$ (dS/AdS-like solutions). The 3rd suggestion is a work in progress since the derived formulas here were only valid for Schwarzschild-like solutions in four dimensions.

\section{Acknowledgements}
R. P. would like to acknowledge networking support of the COST Action CA21106 - COSMIC WISPers in the Dark Universe: Theory, astrophysics and experiments (CosmicWISPers), the COST Action CA22113 - Fundamental challenges in theoretical physics (THEORY-CHALLENGES), and the COST Action CA21136 - Addressing observational tensions in cosmology with systematics and fundamental physics (CosmoVerse).

\bibliography{references.bib}

\end{document}